\documentclass[english,prb]{revtex4}
\usepackage[T1]{fontenc}
\usepackage[latin9]{inputenc}
\usepackage{graphicx}
\usepackage{amssymb}

\usepackage{babel}

\begin{document}

\title{Cold Fermi-gas with long range interaction in a harmonic trap}


\author{Manas Kulkarni$^{1,2}$ and Alexander G. Abanov$^{1}$}

\affiliation{$^{1}$Department of Physics and Astronomy, Stony Brook University,
Stony Brook, NY 11794-3800}

\affiliation{$^{2}$Department of Condensed Matter Physics and Material Science,
Brookhaven National Laboratory, Upton, NY 11973}
\begin{abstract}
We study equilibrium density and spin density profiles for a model of cold one-dimensional spin 1/2 fermions interacting via inverse square interaction and exchange in an external harmonic trap. This model is the well-known
spin-Calogero model (sCM) and its fully nonlinear collective field theory description is known. We extend the field theory description to
the presence of an external harmonic trap and obtain analytic results for statics and dynamics of the system. For instance, we find how the equilibrium density profile
changes upon tuning the interaction strength. The results we obtain for equilibrium configurations are very similar to the ones obtained recently by Ma and Yang \cite{key-2} for a model of fermions with short ranged interactions. Our main approximation is the neglect of the terms of higher order in spatial derivatives in equations of motion -- gradientless approximation \cite{key-1}. Within this approximation the hydrodynamic equations of motion can be written as a set of decoupled forced Riemann-Hopf equations for the dressed Fermi momenta of the
model. This enables us to write analytical solutions for the dynamics
of spin and charge. We describe the time evolution of the charge density when an initial non-equilibrium profile is created by cooling the gas with an additional potential in place and then suddenly removing the potential. 
We present our results as a simple ``single-particle'' evolution in the phase-space reminiscing a similar description of the dynamics of non-interacting one-dimensional fermions.
\end{abstract}
\maketitle
\tableofcontents{}

\section{\label{sec:Introduction}Introduction}

The possibility of creating one and quasi-one-dimensional systems by confining
cold atoms in cigar-shaped harmonic traps\cite{hara-science,key-4,key-5,key-6,wolfgang,d-jin} has raised interest in one-dimensional models of many body systems. These models have been intensively studied since 1970's.
Standard perturbative methods developed in many-body
theory are often not applicable to one-dimensional models because the low dimensionality effectively makes any interaction strong. On the other hand there are some non-perturbative methods which work specifically for particular (integrable) models in one dimension. The 
Bethe Ansatz approach, for example, was successfully used in constructing the complete thermodynamics of quantum integrable systems. However, this approach is  not very suitable for studying the dynamics and correlation functions, due to the complexity of Bethe Ansatz solutions.
To address dynamic questions the method of collective field theory\cite{JevickiSakita,Sakita-book,Jevicki-1992} was developed. It is essentially a hydrodynamic approach to many body systems. 
Many phenomena such as spin-charge dynamics\cite{key-1}, solitons\cite{key-7},
shock waves\cite{key-8}, spin density evolution \cite{key-5,key-6} and sound wave propagation\cite{key-4}
which have become of increasing experimental interest\cite{hara-science,key-4,key-5,key-6,key-9,andrews}
can be studied via the collective field theory formalism.  This formalism can be used both in combination with Bethe Ansatz for integrable models where it can even produce some exact results for dynamical problems or as a phenomenological approach where exact microscopic derivations are too difficult or even impossible. The hydrodynamic approach allows to study truly nonlinear and nonperturbative dynamic behavior\cite{key-1} of many body systems. Linearized hydrodynamic equations are equivalent to the method of bosonization which was widely used for treating interacting systems in one dimension\cite{1981-Haldane_JPC,stone}.

In previous publications\cite{key-1,key-3} the collective field theory approach was applied to the well-known spin-Calogero model\cite{key-11,key-12} to study the coupled nonlinear dynamics of spin and charge. The model has a long range $1/r^{2}$ interaction and is not the easiest one to realize experimentally. On the other hand, the potential $1/r^{2}$ should be considered as relatively short ranged in one dimension.\cite{shranged} The model has another advantage: it is integrable and the integrability is not destroyed by the presence of an external harmonic potential\cite{hikami-harmonic} $V(x)=\frac{m\omega^{2}}{2}x^{2}$. In contrast, for, say, the quantum integrable model of fermions with delta-interaction\cite{1967-Yang} the integrability is destroyed by an external harmonic potential.

In this paper, we explore the effects of the harmonic trap on the collective behavior
of the spin-Calogero model (sCM) at zero temperature. Its Hamiltonian is given by\cite{key-11,key-12,hikami-harmonic,key-10}
\begin{equation}
	H=-\frac{1}{2}\sum_{i=1}^{N}
	\frac{\partial^{2}}{\partial x_{i}^{2}}+\frac{1}{2}
	\sum_{i\neq j}\frac{\lambda(\lambda-P_{ij})}{\left(x_{i}-x_{j}\right)^{2}}
	+\frac{1}{2}\sum_{i=1}^{N}x_{i}^{2}.
 \label{eq:microscopic}
\end{equation}
Here and throughout the paper we take the mass of particles as unity and measure distances in units of an oscillator length $l=\sqrt{\hbar/m\omega}$ and energy in units of $\hbar\omega$. The operator $P_{ij}$ exchanges the positions of particles $i$ and $j$\cite{key-10}. The coupling parameter $\lambda$ is positive and $N$ is the total
number of particles. The last term in (\ref{eq:microscopic}) is the harmonic trap potential. It prevents particles from escaping to infinity and corresponds to effective optical potentials used in experiments to keep particles. The above model (\ref{eq:microscopic}) was shown to be
integrable\cite{hikami-harmonic}. The fully nonlinear hydrodynamics for the
above model (\ref{eq:microscopic}) without an external trap has been
investigated in Ref.~\onlinecite{key-1}. 

The paper is organized as follows. In Sec.~\ref{sec:Collective-Field-theory}
we present the collective description of the microscopic model (\ref{eq:microscopic})
in terms of collective fields and write equations of motion for these fields. 
Then we obtain static solutions: density and spin density profiles in Sec.~\ref{sec:Static-solutions}. We consider the dependence of these equilibrium profiles 
on coupling and find it to be very similar to the recent predictions of Ma and Yang\cite{key-2}
for the model of fermions with contact interaction in a harmonic trap. 
In Sec.~\ref{sec:Dynamics} we show how hydrodynamic fields evolve when the system is
perturbed from the equilibrium configuration. 
We model an initial non-equilibrium profile as the one obtained by cooling
the gas with an additional potential, i.e., keeping what is commonly referred to in literature\cite{key-4} as a ``knife'' - in place (for examples of experiments involving ``knife'' see Refs \onlinecite{andrews} and \onlinecite{key-4}). 
In this section (\ref{sec:Dynamics}) we solve the hydrodynamic equations of motion exactly. These equations written for dressed Fermi momenta are reduced to forced
Riemann-Hopf equations. The solutions of these equations have a very simple form. It turns
out that in the phase-space picture solutions are given just by a rotation by an angle
$t$ ($\omega t$ in physical units) similar to a classical harmonic oscillator. 
We also study the density dynamics and see how an initial density perturbation evolves. The exact  solution of the forced Riemann-Hopf equation is presented in
Appendix \ref{sec:RH_app} for reader's convenience.

\section{\label{sec:Collective-Field-theory}Collective Field theory}

In this section we summarize the main results of the collective
field theory for sCM\cite{key-1,1996-AMY,jevicki-notes} following notations of Ref.~\onlinecite{key-1}. We also include an external harmonic potential which is done in a straightforward way. 
The microscopic Hamiltonian (\ref{eq:microscopic})
is rewritten in terms of hydrodynamic fields: the density of particles
with spin up/down $\rho_{1,2}$ and their respective velocity fields $v_{1,2}$. Although, it is possible to study the exact quantum hydrodynamics of sCM\cite{1996-AMY,jevicki-notes}, in this  paper we neglect hydrodynamic terms with higher order of spatial gradients (gradientless approximation) and treat the equations of motion classically following Ref. \onlinecite{key-1}. This description is referred to as a gradientless
hydrodynamics and is applicable only for sufficiently smooth and slowly evolving
field configurations, where terms with derivatives of fields
can be neglected. A fully nonlinear gradientless hydrodynamics can describe nonlinear phenomena missed in conventional linear bosonization approach. In Ref.~\onlinecite{key-1}, this theory was used to study the
non-linear coupling between the spin and charge degrees of freedom.
\footnote{The sCM is a very special model and it does not exhibit true 
spin-charge separation even in the linear approximation. 
In sCM the spin and charge velocities are the same.} In Ref.~\onlinecite{key-3} the Emptiness Formation probability (a particular $n$-point correlation function) was calculated using instanton approach to the collective field theory of sCM.

The collective field theory for sCM in gradientless 
approximation is remarkably simple and
allows for separation of variables in terms of dressed Fermi momenta\cite{key-1}.
Densities and velocities are expressed as linear combinations of dressed Fermi momenta. 

The aim of this paper is to extend this field theory by including
an external potential, in particular, a harmonic potential. We find analytic solutions for both static profiles and dynamics of charge and spin densities of sCM in the presence of harmonic trap. The non-equilibrium initial configuration is realized by cooling the gas with an appropriate knife in place and then suddenly removing the knife as done in experiments (details are in Sec.~\ref{sec:Static-solutions} and Sec.~\ref{sec:Dynamics}).  For simplicity, we focus here
on a particular hydrodynamics sector of the model, i.e., we assume that the
following inequality is valid at any time everywhere in space
\begin{equation}
	|v_{1}-v_{2}|<\pi(\rho_{1}-\rho_{2}).
 \label{eq:CO_sector}
\end{equation}
In addition, we assume throughout this paper that ``1'' labels the majority spin, i.e., $M_{1}>M_{2}$, where $M_{1,2}$ is the total number of particles with spin 1 and spin 2 respectively, i.e.,
\begin{eqnarray}
	M_{1,2} & = & 
	\int_{-\infty}^{+\infty}\rho_{1,2}\,dx.
 \label{eq:M12def}
\end{eqnarray} 

We introduce particular linear combinations of these fields
\begin{eqnarray}
	k_{R1,L1} & = & 
	v_{1}\pm(\lambda+1)\pi\rho_{1}\pm\lambda\pi\rho_{2},
 \label{eq:k1def} \\
	k_{R2,L2} & = & 
	(\lambda+1)v_{2}-\lambda v_{1}\pm(2\lambda+1)\pi\rho_{2},
 \label{eq:kdef}
\end{eqnarray}
referring to $k_{R,L;1,2}$ as to dressed Fermi momenta (see Ref.~\onlinecite{key-1} for details).
For future convenience we invert (\ref{eq:k1def},\ref{eq:kdef}) to get
\begin{eqnarray}
	\rho_{1} & = & \frac{(k_{R1}-k_{L1})}{2\pi(\lambda+1)}-
\frac{\lambda(k_{R2}-k_{L2})}{2\pi(\lambda+1)(2\lambda+1)},
 \label{eq:rho1h} \\
	v_{1} & = & \frac{k_{L1}+k_{R1}}{2},
 \label{eq:v1h} \\
	\rho_{2} & = & \frac{k_{R2}-k_{L2}}{2\pi(2\lambda+1)},
 \label{eq:rho2h} \\
	v_{2} & = & \frac{k_{L2}+k_{R2}+\lambda(k_{L1}+k_{R1})}{2(\lambda+1)}.
 \label{eq:v2h}
\end{eqnarray}
In terms of dressed momenta in the sector
(\ref{eq:CO_sector}) the hydrodynamic Hamiltonian in harmonic trap in gradientless approximation takes the form
\begin{eqnarray}
	H & = & \frac{1}{12\pi\left(\lambda+1\right)}
	\int_{-\infty}^{+\infty}\left\{ k_{R1}^{3}-k_{L1}^{3}
	+\frac{1}{2\lambda+1}\left(k_{R2}^{3}-k_{L2}^{3}\right)\right\} dx
 \nonumber \\
 	& + & \frac{1}{2}\int_{-\infty}^{+\infty}x^{2}
	\left(\frac{k_{R1}-k_{L1}}{2\pi(\lambda+1)}
	+\frac{k_{R2}-k_{L2}}{2\pi(\lambda+1)(2\lambda+1)}\right)dx.
 \label{eq:collective}
\end{eqnarray}
The first term in (\ref{eq:collective}) was derived in Ref.~\onlinecite{key-1} and the second term is due to the presence of the external harmonic trap.
Notice, that the second term in (\ref{eq:collective}) can be written
as $\int_{-\infty}^{+\infty} \frac{x^{2}}{2}\rho_{c}\,dx$
where $\rho_{c}=\rho_{1}+\rho_{2}$ (\ref{eq:rho1h},\ref{eq:rho2h}) and represents the effect of the harmonic potential.

The Poisson's brackets 
between hydrodynamic fields are given by
\begin{equation}
	\left\{ \rho_{\sigma}(x),v_{\sigma^{\prime}}(y)\right\} 
	=\delta_{\sigma\sigma^{\prime}}\delta^{\prime}(x-y),
 \label{eq:rvcom}
\end{equation}
where $\sigma,\sigma^{\prime}=1,2$ are spin labels.

Then (\ref{eq:rvcom}) along with (\ref{eq:k1def},\ref{eq:kdef})
give the following brackets for $k's$, (here $\alpha,\beta$
take values $R1,\, L1,\, R2,\, L2$)
\begin{equation}
	\left\{ k_{\alpha}(x),k_{\beta}(y)\right\} 
	=2\pi s_{\alpha}\delta_{\alpha\beta}\delta^{\prime}(x-y)
 \label{eq:kcom}
\end{equation}
with 
\begin{eqnarray}
	s_{R1} & = & -s_{L1}=\lambda+1,
 \label{eq:s1}\\
	s_{R2} & = & -s_{L2}=(\lambda+1)(2\lambda+1).
 \label{eq:s2}
\end{eqnarray}

The collective Hamiltonian (\ref{eq:collective}) with Poisson's brackets (\ref{eq:kcom})
generate equations of motion for fields $k_{t}=\left\{H,k\right\}$ which turn out to be the forced Riemann-Hopf equations. Namely, for any $k=k_{1R},\, k_{1L},\, k_{2R},\, k_{2L}$ we have
\begin{equation}
	k_{t}+kk_{x}=-x.
 \label{eq:forced_rh_main}
\end{equation}
For the case of a more general external potential $V(x)$ the right hand side of (\ref{eq:forced_rh_main}) should be replaced by $-\partial_{x}V$. One would expect to arrive at Riemann-Hopf equation \cite{key-1} modified by a force term due to the external potential $V(x)$. It is remarkable, though, that the coupling constant $\lambda$ does not enter (\ref{eq:forced_rh_main}) at all.

\section{\label{sec:Static-solutions}Static solutions}

In this section we consider static density and velocity profiles
of the sCM in harmonic trap in gradientless approximation. 
We give simple analytical expressions
for these profiles for a system with an arbitrary spin polarization having
a fixed number $M_{1}$ of spin-up and $M_{2}$ of spin-down particles.
We describe static profiles in the form of phase-space diagrams.
This description is very simple and gives a direct way to studying the dynamics
of sCM (see Sec.~\ref{sec:Dynamics}).
Although these results are obtained for a particular 1D model with long range interaction (sCM) the static profiles look very similar to the ones recently obtained by Ma and Yang for a fermionic model with contact interaction in harmonic trap\cite{key-2}.
As mentioned earlier this similarity is, probably, due to the fact that 
the potential $1/r^{2}$ can be considered in one dimension as a relatively short ranged.\cite{shranged} 

To obtain equilibrium (static) density and velocity profiles we assume no time dependence in (\ref{eq:forced_rh_main}) and obtain,
\begin{equation}
	\partial_{x}(k^{2}+x^{2})=0\
 \label{dt0}
\end{equation}
which is the equation of a ``circle'' in the phase  space ($x-k$) plane
\begin{equation}
	k^{2}+x^{2}=\mbox{const}.
 \label{eq:circle_eq}
\end{equation}

The constant in (\ref{eq:circle_eq}) depends on $\lambda,\, M_{1}\mbox{ and }M_{2}$ and can be easily determined using (\ref{eq:rho1h},\ref{eq:rho2h}) along with (\ref{eq:M12def}). We find 
that the dressed-momenta for spin-up ($k_{R1,L1}\equiv k_{1}$)
and spin-down ($k_{R2,L2}\equiv k_{2}$) particles satisfy the equations of
two circles respectively. 
\begin{eqnarray}
	k_{1}^{2}+x^{2} 
	& = & 2(\lambda+1)M_{1}+2{\lambda}M_{2},
 \label{eq:k1c}\\
	k_{2}^{2}+x^{2} 
	& = & 2(2\lambda+1)M_{2}.
 \label{eq:k2c}
\end{eqnarray}
The equation (\ref{eq:k1c}) defines a double-valued function $k_{1}(x)$. The positive value is identified with $k_{R1}(x)$ while the negative one gives $k_{L1}(x)$. The values of $k_{R2,L2}(x)$ are obtained in a similar manner from the second equation (\ref{eq:k2c}). 

In order to write the expressions for charge-density ($\rho_{c}=\rho_{1}+\rho_{2}$)
and spin-density ($\rho_{s}=\rho_{1}-\rho_{2}$) it is
convenient to use re-scaled coordinate $\eta$ and re-scaled charge and spin densities $\tilde{\rho}_{c,s}$ given by
\begin{eqnarray}
	\eta & = & \frac{x}{\sqrt{\left(2\lambda+1\right)N}},
 \label{eq:eta}\\
	\tilde{\rho}_{c,s} & = & \sqrt{\frac{\left(2\lambda+1\right)}{N}}\,\rho_{c,s}.
 \label{eq:rtilde}
\end{eqnarray}
Here the total spin $M$, the total number of particles (charge) $N$ and the magnetization $\nu$ are defined respectively as
\begin{eqnarray}
	M & = & M_{1}-M_{2},
 \label{eq:M}\\
	N & = & M_{1}+M_{2},
 \label{eq:N}\\
	\nu & = & \frac{M}{N}.
 \label{eq:xi}
\end{eqnarray}
From (\ref{eq:M}-\ref{eq:xi}) we have
\begin{equation}
M_{1,2}=\frac{N}{2}(1\pm\nu).
\label{eq:M12}
\end{equation}
With these definitions we have from (\ref{eq:k1c},\ref{eq:k2c}) and (\ref{eq:rho1h},\ref{eq:rho2h})
\begin{eqnarray}
	\tilde{\rho}_{c} 
	& = & \frac{\left(2\lambda+1\right)}{\pi(\lambda+1)}
	\sqrt{1+\frac{\nu}{2\lambda+1}-\eta^{2}}
	+\frac{1}{\pi(\lambda+1)}
	\sqrt{1-\nu-\eta^{2}},
 \label{eq:rct}\\
	\tilde{\rho}_{s} 
	& = & \frac{\left(2\lambda+1\right)}{\pi(\lambda+1)}
	\left(\sqrt{1+\frac{\nu}{2\lambda+1}-\eta^{2}}
	-\sqrt{1-\nu-\eta^{2}}\right).
 \label{eq:rst}
\end{eqnarray}

One could also notice as a cross-check that
\begin{eqnarray}
	\int_{-\infty}^{+\infty}\tilde{\rho}_{c}\,d\eta 
	& = & \frac{1}{N}\int_{-\infty}^{+\infty}\rho_{c}\,dx =1,
 \label{eq:rcti}\\
	\int_{-\infty}^{+\infty}\tilde{\rho}_{s}\,d\eta 
	& = & \frac{1}{N}\int_{-\infty}^{+\infty}\rho_{s}\,dx =\nu.
 \label{eq:rsti}
\end{eqnarray}

\begin{figure}
\centerline{\includegraphics[scale=0.8]{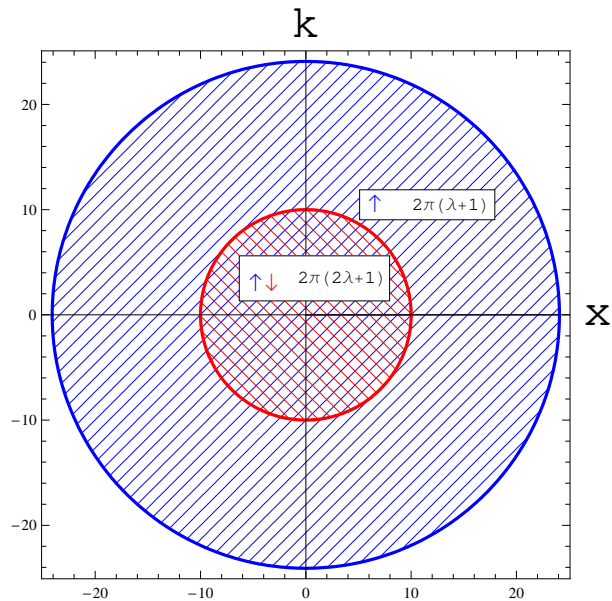}}

\caption{\label{fig:Phase-space-picture-for}Phase-space picture for sCM with $\lambda=2$ in
equilibrium with an overall magnetization $\nu=0.8$. The radii of circles are given by $\sqrt{N(2\lambda+1)(1-\nu)}$ and $\sqrt{N(2\lambda+1+\nu)}$ for inner and outer circles respectively. Particles fill the phase space uniformly with the density of two particles (up and down) per the area $2\pi(2\lambda+1)$ in the inner circle and of one particle (up) per the area $2\pi(\lambda+1)$ in the annulus area between inner and outer circles.}

\end{figure}

\begin{figure}
\centerline{\includegraphics[width=0.6\textwidth]{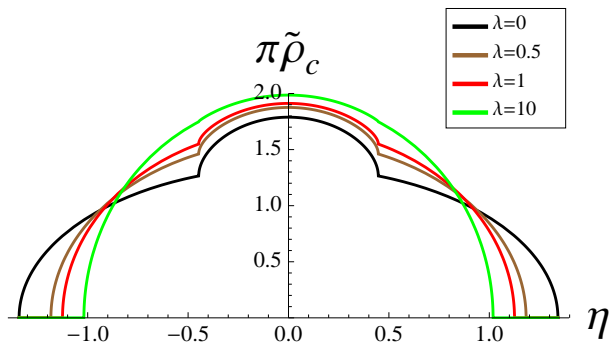}}

\caption{\label{fig:charge}Equilibrium charge density profile for various values of
coupling constant $\lambda$ for fixed magnetization $\nu=0.8$ is shown in re-scaled variables (\ref{eq:eta},\ref{eq:rtilde}). $\lambda=0$ corresponds to noninteracting fermions. Upon increasing the interaction strength $\lambda$ the equilibrium profile
eventually loses its bump feature. This prediction for sCM is very
similar to recent predictions of Ma and Yang for fermions with contact
interaction\cite{key-2}.}

\end{figure}

The analytical expressions for momenta (\ref{eq:k1c},\ref{eq:k2c})
and the analytical expressions for charge density (\ref{eq:rct})
and spin density (\ref{eq:rst}) are the main results of this section. We see that in terms of phase space (Fig.~\ref{fig:Phase-space-picture-for})
the system can be described by two circles of different radii. The radii
depend on the coupling strength $\lambda$ and the numbers of particles
$M_{1}$ and $M_{2}$ as can be seen from (\ref{eq:k1c},\ref{eq:k2c}). 
This is consistent with an exclusion statistics picture\cite{key-1} so that the particle with spin up occupies the area $2\pi(\lambda+1)$ in the phase space while in the domain where both spin up and spin down particles are present the area is $2\pi(2\lambda+1)$ per $2$ particles. In the limit $\lambda=0$ (free fermions) the inner circle in Fig.~\ref{fig:Phase-space-picture-for} is filled with double density compared to the annulus region between the inner and outer circles. This is reflected as a bump feature for the charge density (see Fig.~\ref{fig:charge}). In the limit of a very strong repulsion $\lambda\to +\infty$ the particles are mutually exclusive with the phase-space area approximately $2\pi\lambda$ per particle and the charge density does not show any bump (see Fig.~\ref{fig:charge}). For an arbitrary $\lambda$ the bump feature interpolates between these two limits as shown in Fig.~\ref{fig:charge} with an explicit formula for the static charge density profile given in Eq.~(\ref{eq:rct}). We notice here that the size of the cloud $L_{cigar}=2\sqrt{N(2\lambda+1+\nu)}$ given by the support of (\ref{eq:rct}) grows with $\lambda$ (1D gas expands when $\lambda$ increases). We have taken this main dependence into account by plotting density as a function of re-scaled coordinate $\eta$. 

\begin{figure}
\includegraphics[width=0.6\textwidth]{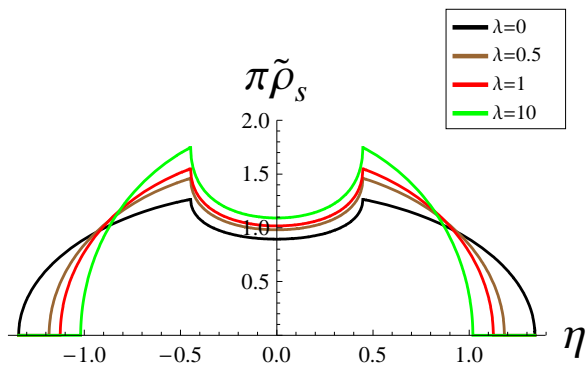}

\caption{\label{fig:spin}Equilibrium spin density profile for various values of coupling
constant $\lambda$ for the fixed magnetization $\nu=0.8$. In re-scaled variables the dip in the spin density profile depends weakly on the interaction strength $\lambda$. Compare with the charge density profile of Fig.~\ref{fig:charge}.  }

\end{figure}

It is interesting to note that although the bump in charge density profile disappears in the limit of a very strong repulsion, the analogous feature in the spin-density profile is present for any $\lambda$. It is practically intact when plotted in terms of re-scaled variables (see Fig.~\ref{fig:spin}) and is given explicitly by (\ref{eq:rst}).

One should expect qualitatively similar profiles for the fermions with short range repulsion in a harmonic trap. Indeed, the physical origin of the bump feature is transparent for noninteracting fermions. It comes just from a superposition of densities of clouds of different size. We expect that the repulsive interaction smears the bump feature making sure that the particles of different species avoid each other similarly to the Pauli exclusion of particles of the same species. Indeed, recent calculations of charge density profiles for fermions with contact interactions by Ma and Yang \cite{key-2} give results very similar (qualitatively) to the ones shown in Fig.~\ref{fig:charge}. It would be interesting to check whether the dip in the spin density profile does not disappear with an increase of the repulsion strength in the model of Ref. \onlinecite{key-2}. 

The following comment is in order. As the origin of the bump feature in the equilibrium charge density profile can be traced to the model of noninteracting fermions, this feature is generic and should be observed also in three-dimensional harmonic traps which are widely used in cold atom experiments. Indeed, the superposition of two ellipsoid-like clouds of spin-up and spin-down particles will give the bump in the overall number density of atoms. The repulsive interaction will smear the feature while the attractive one will amplify it. 

To manipulate cold atom systems experimentally additional external potentials are often used. The gas is cooled in the presence of a harmonic trap and an additional external potential. The latter is usually created by optical means and referred to as ``knife''. It is possible to create an external potential different for different particle species (spin up and spin down here). In the following we concentrate on the potential acting only on ``charge'' degrees of freedom. 

\begin{figure}
\includegraphics[width=0.9\textwidth]{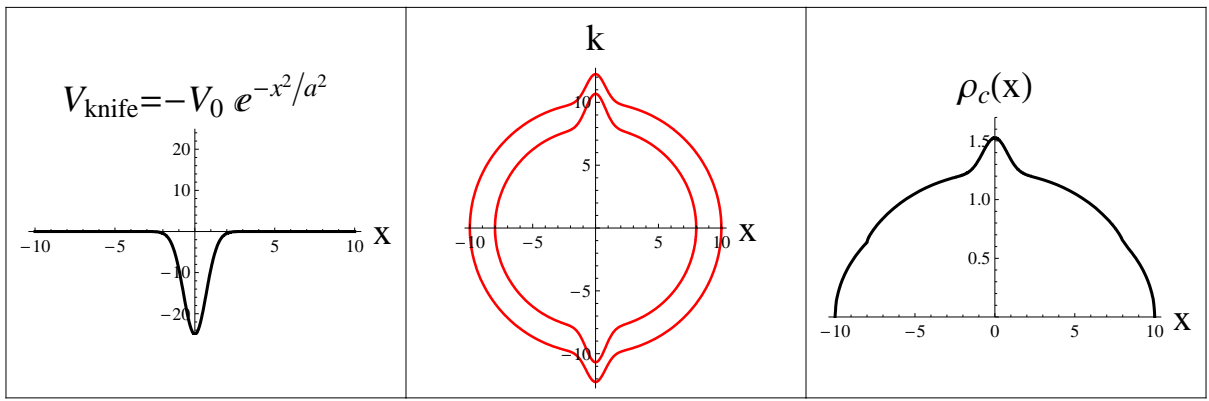}
\caption{\label{fig:statknife}
(left to right) a) Attractive knife potential in the presence of which the fermi gas is cooled. b) The distortion of phase space circles due to the knife. c) The corresponding charge density profile. The values of $\nu$ and $\lambda$ are $\nu=0.5$ and $\lambda=2$.}
\end{figure}

We choose a knife potential of the form $V_{knife}=-V_{0}e^{-x^{2}/a^{2}}$. Then the static equation (\ref{dt0}) is modified as  
\begin{equation}
	\partial_{x}\left(\frac{k^{2}}{2}+\frac{x^{2}}{2}+V_{knife}\right)=0.
 \label{dt0knife}
\end{equation}
Due to the inclusion of the knife (\ref{dt0knife}) the circles in phase-space acquire peaks (attractive knife $V_{0}>0$) or dips (repulsive knife $V_{0}<0$). There is also a peak or dip in the corresponding charge density profile. The knife shape, the corresponding phase-space and the charge density profiles are shown in Fig.~\ref{fig:statknife}. 
After using the knife to create an initial density profile with a peak (dip) one can remove the knife potential and study the evolution of density (or velocity) profiles as a function of time. The initial configuration is a non-equilibrium configuration. Its dynamics is governed by (\ref{eq:forced_rh_main},\ref{eq:k1def},\ref{eq:kdef}). We study the corresponding evolution in the next section Sec.~\ref{sec:Dynamics}. 

\section{\label{sec:Dynamics}Dynamics}

Once the knife potential (Fig.~\ref{fig:statknife}) is suddenly removed we expect the system to evolve. In this section we study this dynamical behavior. Usually, the hydrodynamic equations are coupled partial differential equations. For sCM in gradientless approximation it is possible to separate variables using dressed Fermi momenta instead of charge and spin densities and velocities \cite{key-1}. 
The dressed momenta  are related to hydrodynamic densities and velocities by (\ref{eq:k1def},\ref{eq:kdef}) and satisfy the simple forced Riemann-Hopf equation (\ref{eq:forced_rh_main}). We reproduce it in this section for reader's convenience. 
\begin{equation}
	k_{t}+kk_{x}=-x,
 \label{eq:forced_rh}
\end{equation}
where $k=k_{1R},\, k_{1L},\, k_{2R},\, k_{2L}$.

It is remarkable that the force term (right hand side) is the same for all dressed momenta. This is just another manifestation of the noninteracting nature of Calogero-type models (free particles with exclusion statistics). Of course, physical quantities of interest such as densities
and velocities are some $\lambda$-dependent linear combinations of $k's$ (\ref{eq:rho1h},\ref{eq:v1h},\ref{eq:rho2h},\ref{eq:v2h}) and hence do depend on the coupling strength $\lambda$. 

As it is explained in Appendix \ref{sec:RH_app} the forced Riemann-Hopf equation (\ref{eq:forced_rh}) can be easily solved in parametric form. Given an initial profile $k_{0}(x)=k(x,t=0)$ (specifying the curve in a phase space picture) we can write a profile $k(x,t)$ at time $t$ in a parametric form
\begin{eqnarray}
 	x(s;t) & = & R(s) \sin\left[t+\alpha(s)\right],
 \label{eq:par_x_main}\\
	k(s;t) & = & R(s) \cos\left[t+\alpha(s)\right].
 \label{eq:par_u_main}
\end{eqnarray}
Here the parameter  is $s$ and the functions $R(s)$ and $\alpha(s)$ are determined by an initial profile $k_{0}(x)$ as
\begin{eqnarray}
	\alpha(s) & = & \tan^{-1}\left(\frac{s}{k_{0}(s)}\right),
 \label{eq:theta}\\
	R(s) & = & \sqrt{s^{2}+k_{0}(s)^{2}}
 \label{eq:R_eq}
\end{eqnarray}
consistent with (\ref{eq:par_x_main},\ref{eq:par_u_main}) at $t=0$. 

We immediately notice that the evolution (\ref{eq:par_x_main},\ref{eq:par_u_main}) is just a rotation of the curve $k(x)$ in the $x-k$ phase space with constant angular velocity $1$ ($\omega$ in dimensionfull variables). 

\begin{figure}
\includegraphics[width=0.9\textwidth]{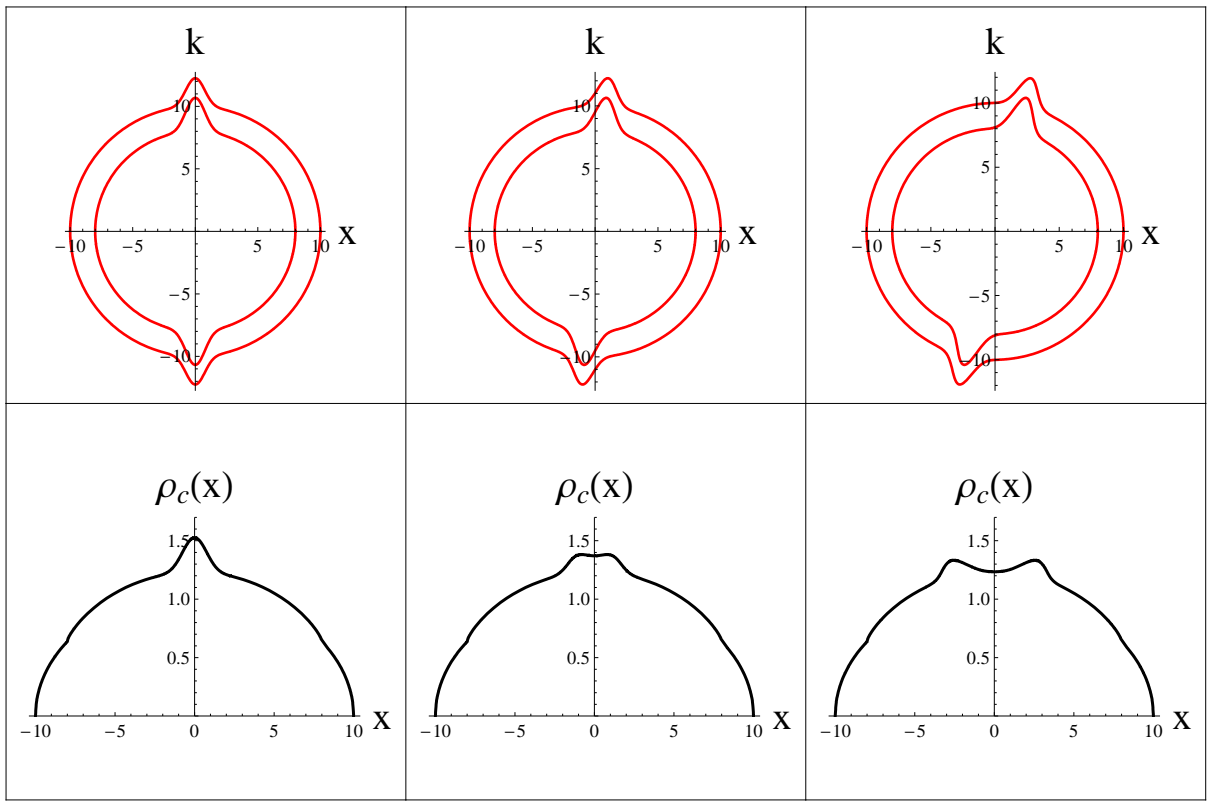}
\caption{\label{fig:Above-rot} 
Top Row: (left to right) Evolution of phase space for time t=0,0.08 and 0.23 respectively. We see that this is merely a rotation by angle $t$.
Bottow Row: Corresponding charge density evolution for times t=0,0.08,0.23. The additional peak created by the attractive knife flattens and eventually splits into two peaks. The values of $\nu$ and $\lambda$ are $\nu=0.5$ and $\lambda=2$.}
\end{figure}

In this section we take the initial profile $k_{0}(x)$ as the one obtained as an equilibrium profile in phase space in the presence of the knife potential (Fig.~\ref{fig:statknife}b). When the knife is removed suddenly this profile serves as an initial non-equilibrium profile. The time evolution in the phase-space is just a rotation of an initial profile by an angle $\theta= t\:$ as shown in Fig.~\ref{fig:Above-rot}.
From the phase space picture at time $t$ obtained by this rotation of an initial profile it is straightforward to compute the charge density evolution using (\ref{eq:rho1h}-\ref{eq:v2h}) (and similarly for spin and velocities evolutions). 

The top row of Fig.~\ref{fig:Above-rot} shows the phase space rotation at various times. The bottom row is the extracted charge density. Since we are dealing with a gradientless theory we can study profile evolution only at times for which the field profiles are smooth. This gradientless approximation is commonly employed in studying nonlinear equations \cite{whitham} and allows to study the evolution for a finite time when the nonlinear terms dominate the terms with higher order in spatial gradients (dispersive terms). Of course, this is possible only if an initial profile is sufficiently smooth.  For longer times, the solution inevitably evolves towards configurations with large field gradients (such as shock waves\cite{key-8,key-9}) and the gradientless approximation becomes inapplicable.
Choosing a sufficiently broad profile of the knife potential we make sure that during the initial stage of the evolution, corrections due to gradient terms in equations of motion are small. We emphasize that this evolution obtained in gradientless approximation already shows some interesting features. 

We see that the central peak (created by cooling with attractive knife) in the charge density profile slowly flattens and eventually splits into two peaks (see the bottom row of Fig.~\ref{fig:Above-rot}). Due to nonlinear effects these two split peaks start to steepen and we expect that at that point, gradient corrections will play a role and the profile would develop dispersive shock waves. 

Although the solution of (\ref{eq:forced_rh}) is not well defined beyond some time (gradient catastrophe time) the parametric solution (\ref{eq:par_x_main},\ref{eq:par_u_main}) can be formally extended beyond that time and produces multiply-valued solutions. These multiply-valued solutions should not be used as the equation (\ref{eq:forced_rh}) has corrections with higher power of gradients which will significantly change the solution beyond the gradient catastrophe time. It is interesting however that in time $t=2\pi$ the solution should reproduce an initial profile and this is an exact feature of the spin-Calogero model. Of course, any corrections to the sCM model destroying integrability will lead to equilibration of the system and the time-periodicity will be lost. 
Probing the charge dynamics in fermi gases at large times is indeed possible experimentally\cite{key-4} and it would be interesting to see this equilibration experimentally.

While our analytical expressions (\ref{eq:par_x_main}-\ref{eq:par_u_main}) are correct for free fermions and sCM (long ranged interactions) we expect qualitatively similar dynamics at small times for systems with short range interaction. The corresponding dynamics studies for fermions with contact interaction (short ranged) could be an interesting extension to the recent
work of Ma and Yang\cite{key-2}. To the best of our knowledge there are no
such dynamic studies for fermions with contact interaction in external
harmonic trap. We would also like to remark that although our predictions for dynamics are for pure 1D systems we do expect a qualitatively similar behavior for the more familiar quasi 1D (cigars) experimental setups\cite{key-4,key-5,key-6}. We leave a quantitative analysis of quasi-one dimensional systems for future work. 

\section{Conclusions}

In this paper we addressed statics and dynamics
of a model of one-dimensional spin 1/2 fermions interacting through a long range inverse square interaction in an external harmonic trap (\ref{eq:microscopic}). While one-dimensional systems with long range interactions are yet to be realized there has been a great progress in experimental studies\cite{key-4,key-5,key-6} of quasi-one-dimensional fermionic models with contact-like interactions in an external harmonic trap. 

Inclusion of an external harmonic trap potential is known to break
integrability of most models (for example models of bosons or fermions with delta interaction). On the contrary, the spin-Calogero model (sCM) (\ref{eq:microscopic}) remains integrable even in the presence of an external harmonic potential \cite{hikami-harmonic}. Similarly, the collective field theory of sCM with harmonic trap retains a rather simple structure with dynamics analogous to the one of non-interacting fermions in harmonic potential. In this paper we used the spin-Calogero model as a toy model of more realistic systems of cold Fermi atoms in quasi-one-dimensional harmonic traps.

Using the collective field theory reviewed in Sec.~\ref{sec:Collective-Field-theory} we studied static density profiles in Sec.~\ref{sec:Static-solutions} as well as dynamics of the model (see Sec.~\ref{sec:Dynamics}). 

The obtained static solutions are found to be qualitatively similar to the static solutions
for fermions with contact interaction in harmonic trap\cite{key-2}. This similarity with short range interactions, probably, can be
attributed to low-dimensionality where $1/r^{2}$ can be considered as
relatively short ranged. The obtained density profiles are given in (\ref{eq:rct},\ref{eq:rst}) and are shown in Figures \ref{fig:Phase-space-picture-for}, \ref{fig:charge} and \ref{fig:spin}. The solution of a similar problem for fermions with contact interactions requires numerical solution of Bethe Ansatz equations in combination with the Thomas-Fermi approximation\cite{key-2}. An interesting feature of an equilibrium solution is the ``bump'' in the charge density profile present for a system with non-zero polarization. The reason for this feature is relatively straightforward as it is already present in the system of non-interacting fermions. It is interesting, however, that the smearing of this bump with an increase of the interaction is qualitatively very similar for both fermions with contact interaction\cite{key-2} and for sCM model (Fig.~\ref{fig:charge}). We also notice that while the bump feature in
charge density (Fig.~\ref{fig:charge})
eventually disappears at strong coupling the spin density profile
remains robust (Fig.~\ref{fig:spin}). It would be interesting to have a calculation of the spin density profile for the contact interaction model considered in Ref.~\onlinecite{key-2}.

To study the dynamics of the cold gas in harmonic trap we create an initial non-equilibrium density profile by ``cooling'' the gas in an additional attractive potential (``knife'').  We choose the form of the attractive knife potential $V_{knife}=-V_{0}e^{-x^{2}/a^{2}}$ similar to the one used in experiments\cite{key-4}. When the knife is suddenly removed the density profile shown in Fig.~\ref{fig:statknife} serves as an initial non-equilibrium profile which is expected to evolve in time.
We show in Sec.~\ref{sec:Dynamics} that the central peak in the charge density profile (created by knife) slowly flattens/broadens and eventually splits
into two peaks (see the bottom row of Fig.~\ref{fig:Above-rot}). Due to nonlinear effects these two split peaks start to steepen and we expect that at that point, gradient
corrections will play a role and the profile would develop dispersive
shock waves\cite{key-8,key-9}. 

In conclusion, we studied equilibrium configurations as well as dynamics of spin-Calogero model in harmonic trap. We argued that the model can serve as a toy model for cold Fermi atoms in one dimensional traps due to the relatively short range nature of an inverse square potential in one dimension. The integrability of spin-Calogero model is not destroyed by the presence of harmonic potential and simple analytic solutions of hydrodynamic equations of motion for this model in gradientless approximation are readily available.

\begin{acknowledgments}
We would like to thank C. N. Yang and Zhong-Qi Ma for discussing with
us the results they obtained for the model of fermions with contact interaction in a harmonic
trap prior to publication. We would also like to thank John Thomas and James Joseph for
several insights into experimental aspects of cold fermions in harmonic
traps. The background for this work is based on our previous work
with Fabio Franchini and we are very grateful to him for several discussions.
A.G.A. was supported by the NSF under the grant DMR-0906866.
\end{acknowledgments}
\appendix

\section{\label{sec:RH_app}Exact solution for Riemann-Hopf equation in arbitrary
potential}

In this appendix we find an exact solution $k(x,t)$ of the forced
Riemann-Hopf equation,
\begin{equation}
	k_{t}+\epsilon'(k) k_{x}=-V^{\prime}(x).
 \label{eq:frh}
\end{equation}
Here $\epsilon(k)$ is some function referred to as ``dispersion'' and the prime means the derivative with respect to the independent variable (e.g., $\epsilon'(k)=\partial \epsilon/\partial k$). The problem is to solve (\ref{eq:frh}) with an initial condition $k=k_{0}(x)$ at $t=0$. This problem can be easily solved using the general method of characteristics \cite{charactersitics}. In this appendix we solve the Euler-type Eq.~(\ref{eq:frh}) rewriting it in the Lagrange formulation. Namely, we notice that the left hand side of (\ref{eq:frh}) can be interpreted as the time derivative in the reference frame of the moving particle. In other words (\ref{eq:frh}) is equivalent to 
\begin{eqnarray}
	\dot{x} &=& \epsilon'(k),
 \label{xmotion} \\
 	\dot{k} &=& -V'(x).
 \label{kmotion}
\end{eqnarray} 
Here dot means time derivative and $\dot{k}(x(t),t) =\partial_{t}k+\dot{x}\,\partial_{x}k$ is nothing else but the left hand side of (\ref{eq:frh}). 

The system of equations (\ref{xmotion},\ref{kmotion}) has a first integral of motion (energy)
\begin{equation}
	\epsilon(k) +V(x) = E,
 \label{energy}
\end{equation}
where $E$ is a time-independent constant determined by an initial condition. We invert (\ref{energy}) and use it and (\ref{xmotion}) to find a solution of (\ref{xmotion},\ref{kmotion}) and, therefore, of (\ref{eq:frh})
\begin{eqnarray}
	k(x) &=& \epsilon^{-1}(E-V(x)),
 \label{kE} \\
 	t &=& \int_{s}^{x}\frac{dy}{\epsilon'(k(y))},
 \label{tk} \\
 	E &=& \epsilon(k_{0}(s)) +V(s),
 \label{Es}
\end{eqnarray}
where the last equation gives the value of the energy $E$ in terms of initial data parametrically given as $x=s$ and $k=k_{0}(s)$. 

The system (\ref{kE}-\ref{Es}) defines the solution $k(x,t)$ of (\ref{eq:frh}) for a given initial profile $k_{0}(x)$. 

Let us now assume that there is no external potential $V(x)=0$ and the dispersion is quadratic $\epsilon(k) = k^{2}/2$. Excluding $E$ and $s$ from (\ref{kE}-\ref{Es}) we have
\begin{equation}
	k=k_{0}(x-kt),
 \label{eq:self_cons}
\end{equation}
which is being solved with respect to $k$ gives a well-known exact solution $k(x,t)$ of (\ref{eq:frh}). 
The solution (\ref{eq:self_cons}) was extensively used in Ref. \onlinecite{key-1} for studies of the dynamics of sCM in the absence of an external potential. 

In the presence of an external harmonic potential $V(x) = \omega^{2}x^{2}/2$ (and $\epsilon(k)=k^{2}/2$) we have from (\ref{kE}-\ref{Es})
\begin{eqnarray}
	k(x) &=& \sqrt{2E-\omega^{2}x^{2}},
 \label{kEharm} \\
 	t &=& \int_{s}^{x}\frac{dy}{\sqrt{2E-\omega^{2}y^{2}}},
 \label{tkharm} \\
 	E &=&\frac{1}{2}\left[k_{0}(s)^{2} +\omega^{2}s^{2}\right].
 \label{Esharm}
\end{eqnarray}
Calculating the integral (\ref{tkharm}) and excluding $E$ from the system (\ref{kEharm}-\ref{Esharm}) we obtain after straightforward manipulations
\begin{eqnarray}
	\omega\,x & = & R(s) \sin\left[\omega t+\alpha(s)\right],
 \label{eq:par_x_main_a}\\
	k & = & R(s) \cos\left[\omega t+\alpha(s)\right],
 \label{eq:par_u_main_a}
\end{eqnarray}
where we introduced
\begin{eqnarray}
	\alpha(s) & = & \tan^{-1}\left(\frac{\omega s}{k_{0}(s)}\right),
 \label{eq:theta_a}\\
	R(s) & = & \sqrt{\omega^{2}s^{2}+k_{0}(s)^{2}}.
 \label{eq:R_eq_a}
\end{eqnarray}

The equations (\ref{eq:par_x_main_a},\ref{eq:par_u_main_a}) together with definitions (\ref{eq:theta_a},\ref{eq:R_eq_a}) give a parametric ($s$ is the running parameter) solution of (\ref{eq:frh}) for the case of quadratic dispersion and harmonic external potential. Putting $\omega\to 1$ in (\ref{eq:par_x_main_a}-\ref{eq:R_eq_a}) we reproduce (\ref{eq:par_x_main}-\ref{eq:R_eq}) used in the main body of the paper.


\end{document}